\documentclass[a4paper, 12pt]{article}

\def\s{\scriptscriptstyle}
\def\b{\begin{eqnarray}}
\def\e{\end{eqnarray}}
\def\1{\hskip1pt}
\def\2{\hskip2pt}
\def\4{\hskip4pt}
\def\5{\hskip5pt}
\def\={\5 = \5}
\def\+{\5 + \5}
\def\-{\5 - \5}
\def\k{\kappa}
\def\nn{\nonumber}
\def\df{\dot{f}}
\def\ddf{\ddot{f}}
\def\dg{\dot{g}}
\def\ddg{\ddot{g}}

\begin{document}

\begin{center}
{\huge\textbf{Bouncing Branes \\}}
\vspace {20mm}
{\large \bf Emil M. Prodanov\footnote{e-mail: prodanov@maths.tcd.ie} \\}
\vspace {3mm}
{\it National Centre for Scientific Research "Demokritos", Athens, Greece} \\
\end{center}
\setcounter{footnote}{0}
\vspace{10mm}

\begin{abstract}
Two classical scalar fields are minimally coupled to gravity in the 
Kachru--Shulz--Silverstein scenario with a rolling fifth radius. A Tolman
wormhole solution is found for a $\mbox{I} \mkern-3.5mu \mbox{R} \times 
\mbox{S}^3$ brane with Lorentz metric and for a $\mbox{I} \mkern-3.5mu 
\mbox{R} \times \mbox{AdS}_3$ brane with positive definite metric.
\vfill
\end{abstract}

\normalsize

\newpage

\section{Introduction}

Two classical scalar fields are minimally coupled to gravity in the 
Kachru--Shulz--Silverstein scenario~\cite{KSS} with a rolling fifth radius. 
One of these fields is time-dependent only, while the other is the dilaton. 
We consider the most general four-dimensional homogeneous and isotropic 
Robertson--Walker metric~\cite{MTW} (with a time-dependent scale factor),
naturally generalized to a five-dimensional Randall--Sundrum~\cite{RS} 
context. We find a Tolman wormhole solution~\cite{Tolman} for a 
$\mbox{I} \mkern-3.5mu \mbox{R} \times \mbox{S}^3$ brane with Lorentz metric
and for $\mbox{I} \mkern-3.5mu \mbox{R} \times \mbox{AdS}_3$ brane with 
positive definite metric. The solution represents a collapsing Universe 
which starts expansion just before encountering a big crunch singularity. 
The Universe reaches a moment of minimal spatial volume. This minimum volume 
edgeless achronal spacelike hypersurface is called a 
{\it bounce}~\cite{Visser}. The wormhole is time-dependent and the bounce 
involves the entire Universe. \\
Our analysis will be along the lines of the one in~\cite{KP}.

\section{The Model}

The action for gravity coupled to two scalars is:
\b
S \ = \int d^{\s 4}x \2 dr \5 (\mathcal{L}^{\s (5)}_{\s MATTER} \+ 
\mathcal{L}^{\s (5)}_{\s GRAVITY}) \2,
\e
where:
\b
\mathcal{L}^{\s (5)}_{\s MATTER} & = &\!\!\!\! - 
\frac{1}{2} \sqrt{|g^{\s (5)}|} 
\nabla^{\mu} \phi \nabla^{\nu} \phi \2 g^{\s (5)}_{\mu\nu} -
\!\sqrt{|g^{\s (5)}|} U(\phi) - \!\sqrt{|g^{\s (4)}|} V(\phi) \delta(r)\2, \\
\mathcal{L}^{\s (5)}_{\s GRAVITY}\!\! & = & \!\!\frac{1}{\k^2} \2 
\sqrt{|g^{\s (5)}|} \2\1 R.
\e
Here, $g^{\s (4)}_{\mu\nu}$ is the pull-back of the five-dimensional metric 
$g^{\s (5)}_{\mu\nu}$ to the domain wall taken to be at $r=0$. The 
wall is represented by a delta-function source with coefficient $V(\phi)$, 
parametrising its tension. 
The scalar $\phi(t, r)$ is:
\b
\phi (t, r) \= \psi(t) \+ \chi(r) \2.
\e
We make the ansatz that both the potentials $U$ and $V$ are of Liouville 
type:
\b
\label{exp1}
U(\phi) & = & U_0 e^{\alpha \phi} \2, \\
\label{exp2}
V(\phi) & = & V_0 e^{\beta \phi} 
\e
and that the metric is: 
\b
\label{metric}
ds^2 = s e^{-A(r)} dt^2 + \frac{e^{-A(r)}g(t)}{\left[1 + \frac{\epsilon}{4}
(x^2 + y^2 + z^2)\right]^2} \2 (dx^2 + dy^2 + dz^2) + q f(t)dr^2 \, . 
\e
This is a natural generalisation of the most general four-dimensional 
homegeneous and isotropic Robertson--Walker metric~\cite{MTW} to a 
five-dimensional Randall--Sundrum context~\cite{RS}. The scale factor 
$g(t)$ is a strictly positive function (we are working with a mostly-plus 
metric) and the function $f(t)$ is strictly positive as well (the metric
is never degenerate).The factors $s$ and $q$ are signs ($s^2 = q^2 = 1$). 
The curvature parameter is $\epsilon = + 1$ (for spherical spatial 
three-sections) or $\epsilon = - 1$ (for hyperbolical spatial 
three-sections). \\
The energy--momentum tensor for the scalar fields is:
\b
T_{\mu\nu} & = & \frac{1}{2} \2 \nabla_{\mu} \phi \2 \nabla_{\nu} 
\phi \2 - \2 \frac{1}{2} \2 g_{\mu\nu}^{\s (5)} \left[\frac{1}{2} \2 
\nabla_{\alpha} \phi \2 \nabla_{\beta} \phi  \2 
g^{\s (5)}\phantom{g}\!\!\!^{\alpha\beta } \2 + \2 U(\phi)\right] \nn \\ \nn \\
&& \hskip120pt - \2\frac{1}{2} \2 \frac{\sqrt{|g^{\s (4)}|}}
{\sqrt{|g^{\s (5)}|}} \2 V(\phi) \2 \delta(r) \2 g^{\s (4)}_{ab} \2 
\delta^a_\mu \2 \delta^b_\nu \2.
\e
Einstein's equations $G_{\mu\nu} = \k^2 T_{\mu\nu}$, equivalently 
written in terms of the Ricci tensor as $R_{\mu\nu} = \k^2 (T_{\mu\nu}
- \frac{1}{3}g^{\s (5)}_{\mu\nu} T^\alpha_{\5 \alpha})$, are:
\b
\label{tt}
- \frac{s}{q}\frac{1}{f} e^{-A} A'^2 + \frac{s}{2q} \frac{1}{f} e^{-A} A'' 
+ \frac{1}{4} \frac{\df^2}{f^2} & & \mkern-25mu -  \frac{1}{2}\frac{\ddf}{f} 
+ \frac{3}{4} \frac{\dg^2}{g^2} - \frac{3}{2} \frac{\ddg}{g} 
\phantom{E \hskip3cm P}  \nn \\
& & \mkern-65mu \= \frac{\k^2}{2} \dot{\psi}^2 +
\frac{\k^2}{3} s e^{-A} U + \frac{\k^2}{6} \frac{s}{f^{1/2}} 
e^{-A} V \delta(r) \, , \\ \nn \\
\label{xx}
2 \epsilon - \frac{1}{q} \frac{g}{f} e^{-A} A'^2 + 
\frac{1}{2q} \frac{g}{f} e^{-A} A'' & & \mkern-25mu
- \frac{1}{4s} \frac{\df}{f} \dg  - \frac{1}{4s} \frac{\dg^2}{g} 
- \frac{1}{2s} \ddg \phantom{very big space} \nn \\
& & \mkern-65mu \= 
\frac{\k^2}{3} e^{-A} g U + \frac{\k^2}{6} e^{-A} \frac{g}{f^{1/2}} 
V \delta(r) \, , \\ \nn \\
\label{rr}
- A'^2 + 2 A'' + \frac{q}{4s} \frac{\df^2}{f} e^A -  & & \mkern-25mu
\frac{q}{2s} \ddf e^A 
- \frac{3q}{4s} \df \frac{\dg}{g} e^A \phantom{very \hskip20pt big space} 
\nn \\ & & \mkern-65mu \= \frac{\k^2}{2} \chi'^2 +
\frac{\k^2}{3} q f U + \frac{2 \k^2}{3} q f^{1/2} V \delta(r) \, ,
\\ \nn \\
\label{tr}
& & \mkern-128mu - \frac{3}{4} A' \frac{\df}{f} \= 
\frac{\k^2}{2} \dot{\psi} \chi' \, .
\e
The $tr$-equation, (\ref{tr}), implies:
\b
\label{psi}
\k \dot{\psi}(t) & = & - \sqrt{3} \5 \frac{\df(t)}{f(t)} \, , \\
\label{chi}
\k \chi'(r) & = & \frac{\sqrt{3}}{2} \5 A'(r) \, .
\e
When $\beta = \frac{\kappa}{2 \sqrt{3}} \, $, the potential $V(\phi)$ can be
written in the form:
\b
\label{v}
V\Bigl(\phi(t,r)\Bigr) \= \frac{1}{\k^2} \frac{1}{q f^{1/2}} W \, ,
\e
where $W$ is a constant. \\
Let us assume that the potential $U(\phi)$ can be written as a function
of $t$ and $r$ in the separable form:
\b
\label{u}
U \Bigl(\phi(t,r)\Bigr) \= \frac{1}{\k^2} \left[ \frac{1}{q f} U_1(r) + 
e^A U_2(t) \right] \, .
\e
At the end we will recast the potential $U$ back into the original 
exponential form (\ref{exp1}). \\
Einstein's equations then reduce to:
\b
\label{1} 
& & \frac{1}{4} \frac{\df^2}{f^2} - \frac{1}{2} \frac{\ddf}{f} +
\frac{1}{4} \frac{\df}{f} \frac{\dg}{g} + \frac{\dg^2}{g^2} 
- \frac{\ddg}{g} - \frac{2 \epsilon s}{g} - \frac{3}{2}
\frac{\df^2}{f^2} \= 0 \, , \\ \nn \\
\label{2}
& & - \frac{1}{4s} \frac{\df}{f} \frac{\dg}{g} - \frac{1}{4s}\frac{\dg^2}{g^2} - \frac{1}{2s} \frac{\ddg}{g} + \frac{2 \epsilon}{g} - \frac{1}{3} U_2 \=
\frac{C}{q f} \, , \\ \nn \\
\label{3} 
& & \frac{1}{2} \frac{\df^2}{f^2} - \frac{3}{2} \frac{\df}{f} \frac{\dg}{g} 
- \frac{\ddf}{f} - \frac{2s}{3} U_2 \= \frac{2sD}{qf} \, , \\ \nn \\
\label{4}
& & A'^2 - \frac{1}{2} A'' + \frac{1}{3} U_1 + \frac{1}{6} W \delta(r)
\= C e^A \, , \\ \nn \\
\label{5}
& & \frac{11}{8} A'^2 - 2 A'' + \frac{1}{3} U_1 + 
\frac{2}{3} W \delta(r) \= D e^A \, , \\ \nn 
\e
where $C$ and $D$ are separation constants. \\
We will be looking for a bounce solution in the form:
\b
\label{solution}
g(t) \= f(t) \= B t^2 + h \5 > \5 0 \, ,
\e
where $B$ is a positive constant, not equal to 1 (so that the pull-back of 
the metric to 4 dimensions is flat only asymptotically\footnote{The 
curvature of the brane is $R = \frac{6 - 6B}{Bt^2 + h}$.}), and $h$ is 
another strictly positive constant. \\
Upon substitution of the solutions (\ref{solution}) into the Einstein's 
equations, (\ref{1}) gives:
\b
B \= -\frac{2 \epsilon s}{3} \, .
\e
On the otherhand, $B$ must be positive. Therefore $B = \frac{2}{3}$ and  
$\epsilon$ and $s$ must have opposite signs. Thus the solution is either a 
brane with spherical three-sections and Lorentz metric or a brane with 
hyperbolical three-sections and positive definite metric. Clearly, these two
solutions can also be related by a Wick rotation (time $t$ is changed to 
$it$ and the positive-curvature spacetime becomes a negative-curvature 
spacetime). \\
Einstein's equations (\ref{2}) and (\ref{3}) are consistent if we choose 
\b
\label{u2}
U_2(t) \= \sigma \, \frac{\dg^2(t)}{g^2(t)} \, ,
\e
where $\sigma$ is a constant. \\
The next Einstein's equation, (\ref{2}), yields that the separation constant
$C$ is $\frac{8 \epsilon q}{3}$ and that $\sigma = -\frac{3}{2 s}$. \\ 
The remaining time-dependent Einstein's equation, (\ref{3}), gives 
$D = \frac{C}{4} = \frac{2 \epsilon q}{3}$. \\
The $r$-dependent Einstein's equations (\ref{4}) and (\ref{5}) yield:
\b
\label{u1}
U_1(r) \= 10 \epsilon q e^{A(r)} - \frac{21}{8} A'(r)^2 \, 
\e
and these two equations reduce to a single equation:
\b
\frac{1}{8} A'^2 - \frac{1}{2} A'' + \frac{1}{6} W \delta(r) \= 
-\frac{2 \epsilon q}{3} e^{A} \, .
\e
A solution of this equation is of KSS~\cite{KSS} type:
\b
\label{A}
A(r) \= \ln \frac{1}{( k |r| + 1)^2} \5 ,
\e
where $k$ is a constant, such that  $k^2 = \frac{4 \epsilon q}{3}$. 
Therefore $\epsilon$ and $q$ must have the same signs ($k^2 = 4/3$). The 
constant $W$ in brane tension $V$ is $-12k$. \\
The equation of motion for the scalar field:
\b
\label{eqm}
\nabla^2 \phi \- \frac{\delta U(\phi)}{\delta \phi} \-
\frac{\sqrt{|g^{\s (4)}|}}{\sqrt{|g^{\s (5)}|}} \2 \frac{\delta V(\phi)}
{\delta \phi} \2 \delta(r) \= 0 \, , 
\e
after integration over the fifth dimension in an infinitesimal interval, 
gives a jump condition acrross the brane:
\b
A'(+0) - A'(-0) \= -4k \, .
\e
Let us now write the potential $\, U \Bigl(\phi(t,r)\Bigr) = 
\frac{1}{\k^2} \left[ \frac{1}{q f} U_1(r) + e^A U_2(t) \right] \,$ back in
the exponential form $\, U = U_0 e^{\alpha \phi} \,$.  
Substituting the solution (\ref{A}) for $A(r)$ into (\ref{u1}) gives:
\b
U_1(r) \= - 4 \epsilon q e^{A(r)} \, .
\e
Using this form of $U_1(r)$, together with (\ref{u2}) for $U_2(t)$ and the
value of $B$, we easily find that:
\b
U \Bigl( \phi(t, r) \Bigr) \= - \frac{4 \epsilon h}{\kappa^2} \, 
\frac{e^A}{g^2} \= U_0 e^{\frac{2 \kappa}{\sqrt{3}} \phi} \, .
\e
For a realistic model, one has to choose $h$ sufficiently small.

\section*{Acknowledgements}
It is a pleasure to thank B. Dolan, V. Gueorguiev, G. Savvidy, S. Sen and
especially C. Kennedy for many useful discussions. Comments from 
C. P. Burgess, S. Giddings, I. Novikov, A. Strominger and N. Tetradis were
also very helpful. The financial support of the EU, grant: 
HPRN--CT--1999--00161, is gratefully acknowledged.

\end{document}